\documentclass[11pt,english,aps,prd,preprintnumbers,tightenlines,
floatfix,nofootinbib,superscriptaddress,fleqn]{revtex4} 

\usepackage{graphicx}
\usepackage{epstopdf}
\usepackage{amsmath,amssymb}
\usepackage{booktabs}
\usepackage{mathrsfs}
\usepackage{multirow}
\usepackage{bm}

\newcommand{\be}{\begin{eqnarray}}
\newcommand{\ee}{\end{eqnarray}}

\newcommand{\ket}{\rangle}
\newcommand{\bra}{\langle}
\newcommand{\del}{\partial}

\def\lsim{\displaystyle\mathop{<}_{\sim}}

\begin{document}
\preprint{INHA-NTG-04/2014}

\title{Pion induced Reactions for Charmed Baryons}

\author{Sang-Ho Kim}
\affiliation{Research Center for Nuclear Physics (RCNP), 
Osaka University, Ibaraki, Osaka, 567-0047, Japan}
\author{Atsushi Hosaka}
\affiliation{Research Center for Nuclear Physics (RCNP), 
Osaka University, Ibaraki, Osaka, 567-0047, Japan}
\affiliation{J-PARC Branch, KEK Theory Center, Institute of Particle and Nuclear Studies,
KEK, Tokai, Ibaraki, 319-1106, Japan}
\author{Hyun-Chul Kim}
\affiliation{Department of Physics, Inha University, Incheon 402-751,
  Republic of Korea}
\author{Hiroyuki Noumi}
\affiliation{Research Center for Nuclear Physics (RCNP), 
Osaka University, Ibaraki, Osaka, 567-0047, Japan}
\author{Kotaro Shirotori}
\affiliation{Research Center for Nuclear Physics (RCNP), 
Osaka University, Ibaraki, Osaka, 567-0047, Japan}

\begin{abstract}%
We study pion induced reactions for charmed baryons $B$, 
$\pi + N \to D^* + B$. 
First we estimate charm production rates in comparison with strangeness production
using a Regge model which is dominated by vector ($D^*$ or $K^*$)  
Reggeon exchange. 
Then we examine the production rates of various charmed baryons $B$ 
in a quark-diquark model.  
We find that the production  of excited states are not necessarily 
suppressed, 
a sharp contrast to strangeness production, which is a unique feature of 
the charm production with a large momentum transfer.  
\end{abstract}

\maketitle

\section{Introduction} 
\label{sec:Introduction}

Observations of new hadrons have been stimulating diverse activities 
in hadron physics, see for instance, Ref.~\cite{Brambilla:2010cs}.
Evidences first observed  at electron facilities such as KEK, SLAC 
and BES~\cite{Choi:2003ue,Aubert:2004ns,Ablikim:2013mio,Liu:2013dau}
are now receiving strong support from recent LHCb experiments~\cite{Aaij:2013zoa,Aaij:2014jqa}.  
Many new hadrons have been found near the threshold regions of charm or 
bottom quarks.  
Intuitively, excited heavy quarks break a string followed by a creation 
of a light quark-antiquark pair, forming the exotic hadrons with multiquarks 
near the threshold.  
To understand the features of the new findings, therefore, requires 
systematic studies of the dynamics from light to heavy quark regions.

So far, many of the new observations were made for mesons.  
In contrast, progress for baryons has not been achieved much.  
In fact, the number of known  heavy quark  baryons is much less 
than that of light quark baryons.  
The study of charmed baryons is important not only for heavy but also for light 
quark dynamics, which in turn will be linked to the physics of the 
new hadrons and eventually to the unsolved problems of QCD.  

Under the above background, an experimental proposal is being made 
for the new pion beam facility at J-PARC~\cite{e50}.  
The expected pion energy will reach over 20 GeV in the laboratory frame 
which is sufficient to excite charmed baryons up to around 1 GeV.   
This is a challenging experiment since there has been no experiment after 
the one at Brookhaven almost thirty years ago~\cite{Christenson:1985ms}.  
The relevant reaction has been chosen, i.e., 
\be
\pi + N \to D^* + B\, , 
\label{basicreaction}
\ee
where $D^*$ is the charmed vector meson and $B$ a charmed baryon.  
The reason $D^*$ is selected in the reaction is due to experimental advantage
as compared to the production of $D$ meson.  

The purpose of this paper is to perform a theoretical study for 
the above reaction, while
experimental feasibility is now under investigation.  
The study of such reactions is a challenging problem, 
because 
1) not many studies have been performed so far, 
2) production rates should reflect structure of charmed baryons, 
and furthermore
3) charm production mechanism from the threshold to a few GeV regions 
is not well understood.  

The structure of charmed baryons have been studied  
in a quark model~\cite{Copley:1979wj,Roberts:2007ni}.
One of  unique features due to the presence of a charm quark 
is the so-called isotope shift.  
In the light flavor sector where the three quarks have 
a similar mass, 
the two independent internal motions of $\rho$ and $\lambda$
modes are degenerate, which in the presence of a heavy quark 
split and appear differently in the spectrum.  
This seems to be the case already in the strange baryons, 
as seen in the inversion of the mass ordering in $\Sigma(1775)$-$\Lambda(1830)$.  
It is then very important to perform systematic studies from 
the light to the heavy flavor sectors.  

This paper is organized as follows.  
In section 2, we estimate the rate of charm production
using a Regge model in comparison with strangeness production.  
In section 3, we compute the production rates of various charmed baryons $B$, 
up to the orbital excitations of $d$-wave ($l=2$) 
in a heavy quark-diquark description of $B$.  
The result indicates that the production of excited states 
$B$ is not necessarily suppressed in comparison with strange 
hyperon production.  
In section 4, we discuss  prospects and summarize the present work.

\section{Estimation of cross sections} 
\label{sec:totalcrosssection}

\begin{figure}[h]
\begin{center}
\includegraphics[width=0.8 \linewidth]{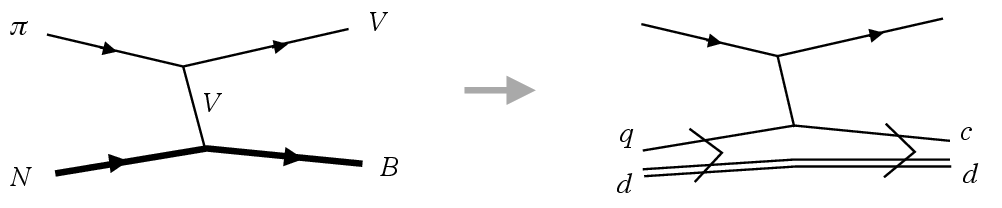}
\end{center}
\vspace{-5mm}
\caption{Left: A $t$-channel process   (vector Reggeon exchange) 
for the $\pi + N \to V + B$ reaction, where $V = D^*, K^*$.  
Right: Quark-diquark structure is shown for the nucleon and charmed baryons, 
which is discussed in section 3.}
\label{fig_t-diagram}
\end{figure}

Let us consider forward angle scattering, 
where the $t$-channel dynamics as shown in Fig.~\ref{fig_t-diagram} 
dominates, and the Regge model  
is expected to be a good prescription.  
Many experiments have shown that  cross sections are of forward peak
(diffractive) at energies beyond a few GeV, which is the region of 
charm production also.  
For strangeness production, a reaction relevant to the present study, 
$\pi + p \to K^* + B_s$, 
was performed long ago ~\cite{Dahl:1967pg,Crennell:1972km}. 
They have shown clearly a forward peak structure, 
which indicates the $t$-channel mechanism in the forward angle region.

%
In the Regge theory~\cite{Donachie2002}, 
the scattering amplitude is first expanded into  
partial waves in the $t$-channel scattering region
($s < 0, t >0$), which is then analytically continued to the physical region of 
$s$-channel scattering  ($s>0, t<0$).  
The sum over integer angular momentum $l$ is then equivalently 
expressed by the Regge pole terms which are the residues 
of the scattering amplitude in the complex angular momentum plane.   
The pole is a function of $t$ and is identified with a Regge trajectory
$\alpha(t)$.  
The amplitude expressed by the Regge poles is then referred to as 
the Reggeon exchange amplitude.  

The advantage of the Regge theory is that it determines 
the asymptotic behavior of the cross section of binary reactions, 
\be
\frac{d\sigma}{dt}  \to s^{2\alpha(t) -2}\, , 
\label{large_s_hehavior}
\ee
which describes well the observed $s$-dependence.  
Among various contributions of different trajectories (Reggeons), 
the dominant one is given by the one of the largest  $\alpha(t)$.  
For example, the vector  Reggeon is more dominant 
than the pseudoscalar Reggeon. 

For our present estimation, we employ the Kaidalov's  prescription
for the vector Reggeon exchange~\cite{Kaidalov:1980bq,Kaidalov:1994mda}, 
\be
\frac{d\sigma}{dt} 
=
\frac{{\rm factor}}{64 \pi |\bm p|^2 s} \Gamma^2(1-\alpha_{V}(t)) 
\left( \frac{s}{\bar s} \right)^2 \left( \frac{s}{s_0} \right)^{2\alpha_{V}(t)-2}\, .
\label{regge_dsdt}
\ee
Here 
$\bm p$ is the relative momentum of the initial state in the center of mass system 
and $\bar s$ a universal scale parameter.  
In the present study of ratios the parameter $\bar s$ is not important.  
The other scale parameter $s_0$ depends on flavors of the Reggeon, and is 
determined by the probabilistic picture~\cite{Kaidalov:1980bq}, 
\be
s_0({\rm charm}) = 4.75 \; {\rm GeV}^2, \; \; \; 
s_0({\rm strange}) = 1.66 \; {\rm GeV}^2\, .
\label{s0}
\ee
For the trajectories $\alpha_{V}(t)$, we employ a non-linear parametrization
\be
\alpha(t) = \alpha_0 + \gamma (\sqrt{T} - \sqrt{T-t})\, , 
\ee
where the parameters $\alpha_0, \gamma$ and $T$ are given in Ref.~\cite{Brisudova:1999ut}.  

In this paper, we show the result of only the differential cross section of Eq.~(\ref{regge_dsdt}).
One could also obtain the total cross section, but here we will not do it, 
because there is ambiguity in the form factor ($t$-dependence).  
In Eq.~(\ref{regge_dsdt}) we employ the one derived from the Regge's method which is 
analytically continued from the $t$-channel scattering region to the $s$-channal 
scattering region.  
This does not necessarily reproduce the observed  $t$-dependence well.
In fact, an alternative parametrization is possible when data are 
available~\cite{Kaidalov:1994mda,Grishina:2005cy,Titov:2008yf}.  
Thus our strategy here is to investigate the forward cross section $d\sigma/dt(\theta = 0)$ 
for charm and strangeness productions, expecting that 
the Regge model works best in the forward angle region.  

\begin{figure}[h]
\begin{center}
\includegraphics[width=0.5 \linewidth]{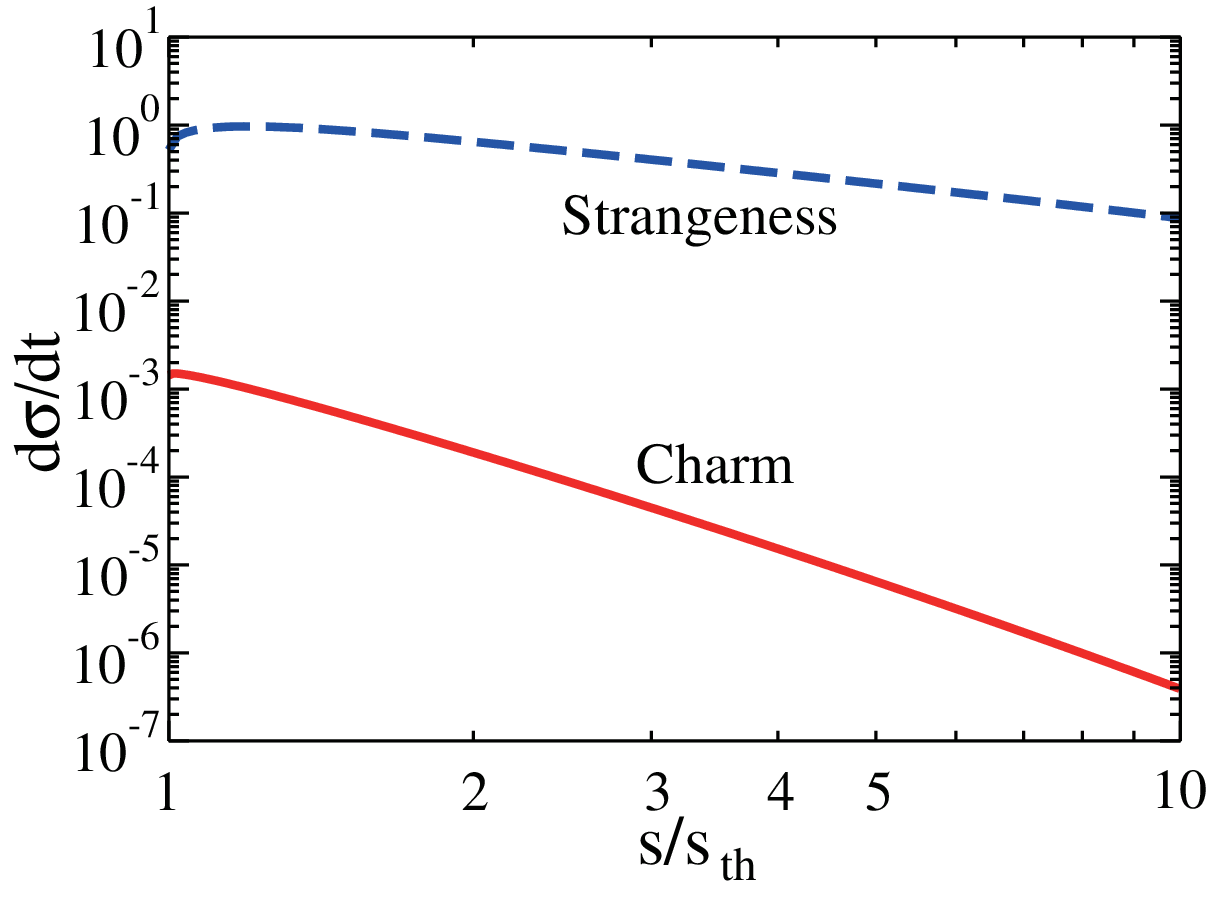}
\end{center}
\vspace{-5mm}
\caption{Forward differential cross sections $d\sigma/dt(\theta = 0)$ as functions of 
$s/s_{\rm th}$, where $s_{\rm th}$ is the $s$-value at the threshold.   
Solid and dashed lines are for charm and strangeness productions, respectively.  
Absolute values are shown in arbitrary unit, but their ratio is properly 
computed by Eq.~(\ref{regge_dsdt}).}
\label{fig_dsdt}
\end{figure}

In Fig.~\ref{fig_dsdt}, we show the results as functions of $s/s_{\rm th}$, 
where $s_{\rm th}$ is the $s$-value at the threshold.  
Two curves are plotted in an arbitrary unit with keeping their ratio
determined by Eq.~(\ref{regge_dsdt}).  
The ratio of the charm to strangeness production 
varies from $10^{-3}$ near the threshold $s/s_{\rm th} \sim 1$ to 
$10^{-5}$ at large energies $s/s_{\rm th} \sim 10$.  
The expected experiments at JPARC will be done most efficiently at 
$s/s_{\rm th} \sim 2$, where the rate of charm production is smaller than 
strangeness production by a factor about $10^{-4}$.  
Therefore, if one uses the $K^*$ production cross sections 
of order $10$ [$\mu$b]~\cite{Dahl:1967pg,Crennell:1972km}, 
the expected one for charm production is of order 1  [nb].  

\section{Production of various charmed baryons}

\subsection{Quark-diquark baryons}

\begin{figure}[h]
\begin{center}
\includegraphics[width=0.25 \linewidth]{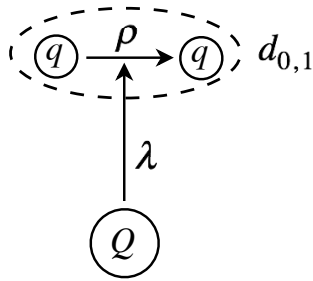}
\end{center}
\vspace{-5mm}
\caption{$\lambda$ and $\rho$ coordinates of a three-quark system, $qqQ$.  
The light quarks $qq$ may form a diquark $d_{S}$ of spin $S = 0, 1$. }
\label{fig_jacobi}
\end{figure}

In this section, baryons are described as two-body systems of a quark and a diquark.
Charmed baryons are then composed of a heavy quark and a light diquark.  
The relative motion of the quark and  diquark is described by the 
$\lambda$ coordinate, one of the Jaccobi coordinates of a three-body system 
as shown in Fig.~\ref{fig_jacobi}.  
The internal motion of the diquark as described by the other variable $\rho$ is 
implicit in the quark-diquark model.  
Due to spin-spin interaction, the pair of $^3S^\rho_0$ quarks ($d^0$) 
is considered to have 
a lower mass than the pair of $^3S^\rho_1$ quarks ($d^1$) .  
In general, we can also consider  internal excitations of diquarks.
Furthermore, the $\lambda$ and $\rho$ modes can couple and mix.  
In this paper, however, we consider only $\lambda$ motions of 
(orbitally) ground state diquarks of the above two kinds, $d^0$ and $d^1$, 
because the reaction mechanism that we consider as shown in Fig.~\ref{fig_t-diagram} (right) 
excites dominantly a $\lambda$ mode.  
The quark-diquark wave functions of the $\lambda$ modes 
are summarized in Appendix B.  
We have then made a tentative assignment of these states with
the nominal ones listed in PDG when available~\cite{Beringer:1900zz}
as shown in Table~\ref{C_and _R}.  
We have also made arbitrary assignment for the unknown states 
to fill the corresponding ones by simply guessing their masses.  
The latter are shown in Table~\ref{C_and _R} with a $^*$ symbol.  

As shown in Fig.~\ref{fig_t-diagram}  in the $t$-channel process,  
a charmed Reggeon is exchanged and couples with a quark in the initial nucleon 
transformed into a charm quark forming a charmed baryon in the final state.  
Our calculation here is performed under several assumptions.

\begin{itemize}

\item
As in the previous section, 
we consider vector ($V = D^*$ or $K^*$) Reggeon exchanges because 
at high energies  the $V$ Reggeon dominates.  

\item
The cross section shows a forward peak.  
Therefore, we compute the differential cross sections only at the 
forward angle.  

\item 
We focus on ratios
of excited charmed baryon production as compared to  
ground state production.  

\end{itemize}

The main issue in this section is the computation of various baryon matrix elements, 
which determines the production rates.  
For this purpose, we  need a vertex for quark-diquark baryons.  
In addition, we also consider a meson vertex to evaluate the whole $t$-channel diagram.  
Thus we introduce the following two interaction Lagrangians, 
\be
{\cal L}_{\pi VV} &=& f  \epsilon_{\mu \nu \alpha \beta}\
\del^\mu \pi \del^\nu V^{\alpha} V^{\beta}\, , 
\label{L_piVV}\\
{\cal L}_{Vqc} &=&g \bar c \gamma^\mu q  V_\mu\, .  
\label{L_Vqc}
\ee
Here,  $f$ and $g$ are coupling constants, 
and $q$ and $c$ denote the spinors of the light ($q = u, d$) and charm quarks, respectively.  

\subsection{Amplitudes}

Let us first look at the matrix element of the $\pi VV$ coupling of Eq.~(\ref{L_piVV}), 
\be
\bra V(k_V) | {\cal L}_{\pi VV} | \pi(k_\pi) V(q) \ket
\sim 
2 f \epsilon_{\mu 0 \alpha \beta}
k_\pi^\mu k_{V}^0 e^\alpha e^\beta
\to
2f k_{V}^0 \vec k_\pi \times \vec e \cdot \vec e\, , 
\label{L_pVV_NR}
\ee
where $k_\pi, k_{V}$ and $q$ are the momentum of 
the initial pion, of the final $V$ and of the exchanged $V$ meson, respectively.
$e^{\alpha, \beta}$ are the polarization vectors of either the final 
or the intermediate vector mesons.  
In these manipulations, we selected the dominant term
assuming that the reaction energy is not relativistically too large
as in the case for $s/s_0 \lsim 2$.  

Next, we compute the baryon matrix element of ${\cal L}_{Vqc}$, 
\be
\bra {\cal L}_{Vqc}\ket &=& \bra g \bar c \gamma^\mu V_\mu q \ket
\nonumber \\
&=& g \varphi_f^\dagger 
\left(1, - \frac{\vec \sigma \cdot \vec p_f}
       {m_c + E_c} \right)
\left(
\begin{array}{cc}
V^0 & - \vec \sigma \cdot \vec V\\
\vec \sigma \cdot \vec V & - V^0 
\end{array}
\right)
\left(
\begin{array}{c}
1\\
\displaystyle{ \frac{\vec \sigma \cdot \vec p_i} {m_q + E_q}}
\end{array}
\right)
\varphi_i\, , 
\ee
where $\varphi_{i, f}$ are the two component spinors 
for the initial light quark and the final charm quark, respectively.  
To proceed, we pick up only terms that contain the spatial component 
of the $V$ meson, because when this $V$ meson is contracted 
with another from the $\pi VV$ vertex, only the spatial 
component survives as Eq.~(\ref{L_pVV_NR}) implies.
Hence we find
\be
{\cal L}_{Vqc} \sim
- g \varphi_f^\dagger
\left[
\left(
\frac{\vec p_f}{m_c+E_c} + \frac{\vec p_i}{m_q+E_q}
\right)
\cdot \vec V
+
i \vec \sigma \times 
\left(
\frac{\vec p_f}{m_c+E_c} - \frac{\vec p_i}{m_q+E_q}
\right)
\cdot
\vec V
\right]
\varphi_i \, .  
\label{L_VNB_NR}
\ee

Now combining the  matrix elements 
Eqs.~(\ref{L_pVV_NR}) and (\ref{L_VNB_NR}), we can write down 
the scattering amplitude as
\be
t_{fi} \sim 
2fg k_{V}^0 \vec k_\pi \times \vec e \cdot 
\vec J_{fi} 
G_{V}(t) \, , 
\label{scattering_t}
\ee
where 
\be
G_{V}(t) 
= \Gamma(1-\alpha(t)_{V}) 
\left( \frac{s}{s_0} \right)^{\alpha(t)_{V}-1}
\label{Gt}
\ee
 is the Reggeon propagator, and 
$\vec J_{fi}$ the baryon transition current,   
\be
\vec J_{fi} 
=
\int d^3x \, \varphi_f^\dagger
\left[
\frac{\vec p_f}{m_c+E_c} + \frac{\vec p_i}{m_q+E_q}
+
i \vec \sigma \times 
\left(
\frac{\vec p_f}{m_c+E_c} - \frac{\vec p_i}{m_q+E_q}
\right)
\right]
\varphi_i
\, e^{i \vec q_{eff} \cdot \vec x} \, .  
\label{Jfi}
\ee
Here we have defined the effective momentum transfer 
\be
\vec q_{eff} = \frac{m_d}{m_d + m_q} \vec P_N
- \frac{m_d}{m_d + m_c}\vec P_B
\label{qeff}
\ee
which takes into account the recoil of the center of mass motion due to the change 
in the masses of $q$ and $c$ quarks~\cite{Itonaga:1990iq}.  

To further simplify the computation, the quark momenta
$\vec p_i$ and $\vec p_f$ are approximated to take 
a fraction of the baryon momentum, 
\be
\vec p_i &\sim& \frac{1}{3} \vec P_N, \; \; \; 
\nonumber \\
\vec p_f &\sim& \frac{m_c}{m_c+m_d} \vec P_B\, .
\label{fraction_pi_pf}
\ee
Note that for the initial state the pion momentum (and hence the nucleon momentum)
is sufficiently large such that the mass of the light quarks in the nucleon is neglected.  
Now for forward scattering where all momenta are collinear along the $z$-axis, 
only the spin current term survives in the scattering amplitude:
\be
t_{fi}
&\sim& 
\left(
\frac{P_B}{2(m_c+m_d)} - 1 
\right)\, 
k_{V}^0 \vec k_\pi \times \vec e \cdot 
\bra f | \,
\vec \sigma \times 
\hat z
\, e^{i \vec q_{eff} \cdot \vec x}\,
| i \ket 
\; G_{V}(t)
\nonumber \\
&=&
\left(
\frac{P_B}{2(m_c+m_d)} - 1 
\right)\, 
k_{V}^0 
\bra f | \,
\left(
(\vec k_\pi \cdot \vec \sigma) (\vec e \cdot  \hat z)
-
(\vec k_\pi \cdot \hat z) (\vec e \cdot  \vec \sigma)
\right)
\, e^{i \vec q_{eff} \cdot \vec x}\,
| i \ket 
\; G_{V}(t)\; , 
\ee
where the constant factors which are irrelevant when taking ratios
of the production rates are ignored.   
The polarization of $V$ can be either  longitudinal ($z$) or transverse ($x, y$), 
but the longitudinal contribution vanishes.  
Moreover, for the transverse polarization, the first term vanishes.  
Finally, we obtain a rather concise formula for the amplitude
\be
t_{fi}
&\sim& 
\left(
\frac{P_B}{2(m_c+m_d)} - 1 
\right)
k_{V}^0 k_\pi
\bra f | \,
\vec e_\perp \cdot  \vec \sigma
\, e^{i \vec q_{eff} \cdot \vec x}\,
| i \ket 
\; G_{V}(t)\, . 
\ee
Here $\vec e_\perp$ denotes the transverse vector, 
and hence the transverse spin induces the transition, 
as expected for the vector ($J^P = 1^-$) exchange process.  

\subsection{Production rates}

We have computed the transition amplitudes $t_{fi}$ from the nucleon $i \sim N$ 
to various charmed baryons $f \sim B$.  
For charmed baryons, we consider all possible states 
including the ground, $p$-wave and $d$-wave excitations.    
The production rates are computed by
\be
{\cal R} \sim  \frac{1}{\rm Flux} \times \sum_{fi} |t_{fi}|^2 \times  {\rm Phase \; space}.
\ee
Using the results of the amplitudes as shown in Appendix A, we find 
\be
{\cal R} (B(J^P)) =
\frac{1}{4 |\bm p| \sqrt{s}}
\gamma^2 \, K^2\, C\,   | I_L |^2 
\frac{q}{4\pi \sqrt{s}}\, .
\label{formula_R}
\ee
In these expressions, $C$ is the geometric factor of the matrix element
$
\bra f | \,
\vec e_\perp \cdot  \vec \sigma
\, e^{i \vec q_{eff} \cdot \vec x}\,
| i \ket 
$
determined by the spin, angular momentum and total spin of the baryon, 
while $I_L  (L = 0, 1, 2)$ contains dynamical information of the baryon wave function.  
$K$ is the  kinematic factor
\be
K = 
k_{V}^0 k_\pi
\left(
\frac{P_B}{2(m_c+m_d)} - 1 
\right)
\;
G_{V}(t)
\ee
and $\gamma$ the following isospin overlap factor
\be
\gamma &=& \frac{1}{\sqrt{2}} \; \;\; {\rm for} \; \Lambda \; {\rm baryons} \, , 
\nonumber \\
&=& \frac{1}{\sqrt{6}}\; \;\; {\rm for} \; \Sigma \; {\rm baryons}\, , 
\ee

\begin{table}[htdp]
\caption{Baryon masses $M$ [MeV] (see text for assignment), 
spin-dependent coefficients $C$ and the ratios of production rates
${\cal R}$ given in Eq.~(\ref{formula_R}).
The second and third rows are the ratios ${\cal R}$ for the strange and charmed baryons, respectively, 
which are normalized to the ground state $\Lambda$.
They are computed 
at $k_\pi^{Lab} = 4.2$ GeV for the strange, and 
at $k_\pi^{Lab} = 20$ GeV for the charmed baryons.}
\begin{center}
\begin{tabular}{c c c c c c c c c }
\hline
\hline
$l =0$ & 
$\Lambda(\frac{1}{2}^+)$ & 
$\Sigma(\frac{1}{2}^+)$ & 
$\Sigma(\frac{3}{2}^+)$ & 
 & & & &  \\
 $M$ [MeV]  & 1116 & 1192 & 1385 &  &  &  &  &   \\
    & 2286 & 2455 & 2520 &  &  &  &  &   \\
\hline
$C$  & 1 & 1/9 & 8/9 & & & & &   \\
${\cal R}(B_s)$  & 1 & 0.04 & 0.210 & & & & &   \\
${\cal R}(B_c)$  & 1 & 0.03 & 0.17 & & & & &   \\
 \hline
\hline
$l = 1$ & 
$\Lambda(\frac{1}{2}^-)$ & 
$\Lambda(\frac{3}{2}^-)$ & 
$\Sigma(\frac{1}{2}^-)$ & 
$\Sigma(\frac{3}{2}^-)$ & 
$\Sigma^\prime(\frac{1}{2}^-)$ & 
$\Sigma^\prime(\frac{3}{2}^-)$& 
$\Sigma^\prime(\frac{5}{2}^-)$  & \\
$M$ [MeV] & 1405 & 1520 & 1670 & 1690 & 1750 & 1750 & 1775 & \\
     & 2595 & 2625 & 2750 & 2800 & 2750 & 2820 & 2820 & \\
\hline
$C$ & 1/3 & 2/3 & 1/27 & 2/27 & 2/27 & 56/135 & 2/5 & \\
${\cal R}(B_s)$ & 0.07 & 0.11 & 0.002 & 0.003 & 0.003 & 0.01 & 0.01 & \\
${\cal R}(B_c)$ & 0.93 & 1.75 & 0.02 & 0.04 & 0.05 & 0.21 & 0.21 & \\
\hline
\hline
$l = 2$ & 
$\Lambda(\frac{3}{2}^+)$ & 
$\Lambda(\frac{5}{2}^+)$ & 
$\Sigma(\frac{3}{2}^+)$ & 
$\Sigma(\frac{5}{2}^+)$ & 
$\Sigma^\prime(\frac{1}{2}^+)$ & 
$\Sigma^\prime(\frac{3}{2}^+)$& 
$\Sigma^\prime(\frac{5}{2}^+)$ &
$\Sigma^\prime(\frac{7}{2}^+)$ \\
$M$ [MeV] & 1890 & 1820 & 1840 & 1915 & 1880 & 2000$^*$ & 2000$^*$ & 2000$^*$ \\
      & 2940 & 2880 & 1840 & 3000$^*$ & 3000$^*$ & 3000$^*$ & 3000$^*$ & 3000$^*$ \\
\hline
$C$ & 2/5 & 3/5 & 2/45 & 3/45 & 2/45 & 8/45 & 38/105 & 32/105 \\
${\cal R}(B_s)$ & 0.02 & 0.04 & 0.003 & 0.001 & 0.001 & 0.001 & 0.001 & 0.001 \\
${\cal R}(B_c)$ & 0.49 & 0.86 & 0.01 & 0.02 & 0.01 & 0.05 & 0.11 & 0.09 \\
\hline
\end{tabular}
\end{center}
\label{C_and _R}
\end{table}%

By using the baryon wave functions as summarized in Appendix B and C, 
the geometric factors $C$ and the production rates ${\cal R}$ are computed.  
In Table \ref{C_and _R}, results are shown for both charm and strangeness productions at 
the pion momentum in the laboratory frame, 
$k_\pi^{Lab} = 20$ GeV for charm production and 
$k_\pi^{Lab} = 4.2$ GeV for strangeness production.  
These momenta correspond to $s/s_{\rm th} = 2$ for both cases.  
The wave functions of strange baryons are obtained by replacing the charm quark 
by a strange quark.  
The rates ${\cal R}$ presented in the table 
are normalized by that of the lowest $\Lambda$ baryon.  

Herein below we make several observations.  

\begin{itemize}

\item
In general the production rates for $\Lambda$ baryons 
are larger than for $\Sigma$ baryons.  
This is a consequence of  SU(6) symmetry of the quark-diquark baryons.  

\item
Some excited $\Lambda_c$ states with a higher $l$ have a similar or even larger production rate
than the ground state, in particular $\Lambda_c(1/2^-)$ and $\Lambda_c(3/2^-)$, 
and  $\Lambda_c(3/2^+)$ and $\Lambda_c(5/2^+)$.  
This is due to large overlap of the wave functions when 
the momentum transfer is large, typically around 1 GeV for charm production.   
The momentum transfer value together with the size of the baryons determines
an optimal angular momentum transfer $\Delta l$.  
For charm production this occurs at around $\Delta l \sim 1$, while for strangeness 
production at $\Delta l << 1$.  
Mathematically, this is explained by the combination
of the power term $(q_{eff}/A)^l$ and the form factor 
$\exp(-(q_{eff}/2A)^2)$ as in Eqs.~(\ref{I_1}) and (\ref{I_2}).  
In hypernucleus production, the same mechanism has been well appreciated, 
demonstrating the success in the studies of reaction and structure~\cite{Itonaga:1990iq}. 

\item 
The above pairs of $\Lambda$'s form a spin-orbit ($LS$) doublet in the quark model, 
or in the heavy quark limit the heavy quark doublet~\cite{Yamaguchi:2014era}.  
Their relative production rates are then determined in a model-independent manner 
up to a kinematic factor.  

\item
We can similarly compute the amplitude for $P$(pseudoscalar)-Reggeon exchanges, 
by replacing the transverse spin by the longitudinal spin, 
$\vec e_\perp \cdot  \vec \sigma \to \vec e_{||} \cdot  \vec \sigma$.  
Although we do not consider this process in this paper, 
a unique feature is that $V$ and $P$ Reggeon exchanges 
do not interfere in the forward amplitude due to the spin selection rule.  

\item
So far, we have looked at $V$ (= $D^*$ or $K^*$) meson production due to the planned 
experimental requirements.  
Theoretically, we can also study the reactions followed by $D$ or $K$ meson production.  
In this case, pseeudoscalar and scalar exchanges are possible, for which we can write down 
similar formulas.  

\end{itemize}

\section{Discussions and Remarks}
                          
We have studied charm production induced by the high-moment pion beam.  
This is a very challenging problem since no experiment has been performed for 
almost thirty years after the one at Brookhaven~\cite{Christenson:1985ms}.  
However, charmed baryon spectroscopy will bring 
us with fruitful information for yet unexplored region in hadron physics.  
This has been the primary motivation of the present study. 

We have first estimated that in the Regge model charm production is 
suppressed by a factor $10^{-4}$ as compared to strangeness production, 
implying an expected cross section of order 1 [nb].  
Another yet important finding in the present study 
is that the production rates of excited charmed baryons
are not necessarily suppressed as compared to those of the ground state.  
This is a consequence of good overlaps of the initial and final state baryons
at the momentum transfer around 1 GeV, 
providing us with more opportunity for the study of excited states.  

In the present study, we have used a simple quark and diquark model for baryons.  
In view of the successes of the constituent picture for low lying states, we expect
some of the features should persist in the charm production reactions also.  
In particular, the identification of $\lambda$ and $\rho$ 
modes should be very important to reveal the mechanism of hadron excitations.  
Further investigations for productions and decays in the heavy quark region 
may provide good information of it.  

\section*{Acknowledgements:}

We thank A.I. Titov, M. Oka, K. Sadato and T. Yoshida for discussions.  
This work is supported in part by the Grant-in-Aid  for Science Research (C) 26400273.  
S.H.K. is supported by Scholarship of the Ministry of 
Education, Culture, Science and Technology of Japan.
The work of H.-Ch.K. was supported by Basic Science
Research Program through the National Research Foundation of Korea
funded by the Ministry of Education, Science and Technology (Grant
Number: 2013S1A2A2035612).

\appendix

\section{Matrix elements}

Let us calculate the matrix elements 
$
\bra f | \,
\vec e_\perp \cdot  \vec \sigma
\, e^{i \vec q_{eff} \cdot \vec x}\,
| i \ket
$
for baryons $B$ with various spin and parity $J^P$.  
For forward scattering, due to helicity conservation, it is sufficient to consider 
only one helicity flip transition for a given $J$
(remember that  only transverse polarization transfer is possible), 
\be
i \to f = J_z(N)  \to (J_z(B), h)  = 1/2 \to (-1/2, 1)
\ee
for $J = 1/2$ and 3/2, 
and 
\be
J_z(N) \to (J_z(B), h ) =  -1/2 \to (-3/2, 1)
\ee
for 
 $J = 3/2$.
 Here $h$ denotes the helicity of the vector meson $V$.  
 Other amplitudes are related to these elements by time reversal.  
 
 The total cross section is then proportional to the sum of 
 squared amplitudes over possible spin states.
 For $J = 1/2$
\be
\sigma &\sim& 
|\bra -1/2, +1| t | + 1/2 \ket |^2 + 
|\bra +1/2, -1| t |  -1/2 \ket |^2
\nonumber \\
&=& 
2 |\bra -1/2, +1| t | + 1/2 \ket |^2
\ee
and for $J = 3/2$ and $5/2$
\be
\sigma &\sim& 
2 \left(
|\bra -1/2, +1| t | + 1/2 \ket |^2 + 
|\bra +3/2, -1| t |  +1/2 \ket |^2
\right) \, .
\ee

\subsection{$N(1/2^+) \to$ ground state baryons}

First we consider the transition to $\Lambda(1/2^+)$ (of both charm and strangeness) 
\be
\bra 
\psi_{000} \chi^\rho _{-1/2}V(+1) |
\vec e_\perp \cdot  \vec \sigma
\, e^{i \vec q_{eff} \cdot \vec x}\,
| \psi_{000} \chi^\rho_{+1/2} \ket\, , 
\ee
where the baryon orbital wave functions $\psi_{nlm}$ are given in Appendix C.  
Note that since the diquark behaves as a spectator 
in the reaction (Fig.~\ref{fig_t-diagram}), 
the good diquark component of $\chi^\rho$ for the nucleon is taken.  
The spectroscopic (overlap) factor of the good diquark component 
in the nucleon is tabulated in below where isospin factor is included also.  
Choosing the $V$ polarization as $\vec e_\perp$, 
we have 
\be
\bra 
\psi_{000} \chi^\rho _{-1/2} |
\sqrt{2} \sigma_-
\, e^{i \vec q_{eff} \cdot \vec x}\,
| \psi_{000} \chi^\rho_{+1/2} \ket
=
\bra \chi^\rho _{-1/2} | \sigma_- |\chi^\rho_{+1/2}\ket \, 
\bra \psi_{000} | \sqrt{2}  \, e^{i \vec q_{eff} \cdot \vec x}\,
|\psi_{000} \ket\, , 
\ee
where the spin and orbital parts are separated and 
$\sigma_-$ is the spin lowering matrix given as 
\be
\sigma_- =
\left(
\begin{array}{cc}
0 & 0 \\
1 & 0
\end{array}
\right) \, . 
\ee
The spin matrix elements are easily computed as 
\be
\bra \chi^\rho _{-1/2} | \sigma_- | \chi^\rho_{+1/2}\ket &=& 1\, , 
\nonumber \\
\bra \chi^\lambda _{-1/2} | \sigma_- | \chi^\lambda{+1/2}\ket &=& - \frac{1}{3}\, , 
\nonumber \\
\bra \chi^S _{-1/2} | \sigma_- | \chi^\lambda{+1/2}\ket &=& \frac{\sqrt{2}}{3}\, , 
\nonumber \\
\bra \chi^S _{-3/2} | \sigma_- | \chi^\lambda{-1/2}\ket &=& - \sqrt{\frac{2}{3}}\, , 
\label{spin_matrix_element}
\ee
where we have shown all relevant matrix elements in the following calculations.  
Therefore, the remaining is the elementary integral over the radial 
distance $r$ with Gaussian functions.  
We find
\be
\Lambda(1/2^+): \; \; 
\bra 
\psi_{000} \chi^\rho _{-1/2} |
\sqrt{2} \sigma_-
\, e^{i \vec q_{eff} \cdot \vec x}\,
| \psi_{000} \chi^\rho_{+1/2} \ket
&=& I_0 \, , 
\ee
where the radial integral $I_0$ is given by 
\be
I_0 
&=& 
\bra 
\psi_{000} |
\sqrt{2} 
\, e^{i \vec q_{eff} \cdot \vec x}\,
| \psi_{000} \ket
=
\sqrt{2} \, \left(
\frac{\alpha^\prime \alpha}{A^2}
\right)^{3/2}
e^{-q_{eff}^2/(4A^2)} \, , 
\nonumber \\
& & A^2 =  \frac{\alpha^2 + \alpha^{\prime 2}}{2}\, . 
\ee
The oscillator parameters are 
$\alpha$ and $\alpha^\prime$ are for the initial and final state baryons, respectively.  

Similarly, we calculate the transitions to the ground state $\Sigma$'s, 
picking up the $\chi^\lambda$ part for the nucleon wave function.  
Only the difference is the spin matrix element which are computed 
by making Clebsh-Gordan decompositions.  
Results are  
\be
\Sigma(1/2^+): \; \; 
\bra 
\psi_{000} \chi^\lambda _{-1/2} |
\sqrt{2} \sigma_-
\, e^{i \vec q_{eff} \cdot \vec x}\,
| \psi_{000} \chi^\lambda_{+1/2} \ket
&=&
-\frac{1}{3} I_0 \, , 
\nonumber \\
\Sigma(3/2^+): \; \; 
\bra 
\psi_{000} \chi^S _{-1/2} |
\sqrt{2} \sigma_-
\, e^{i \vec q_{eff} \cdot \vec x}\,
| \psi_{000} \chi^\lambda_{+1/2} \ket
&=&
\frac{\sqrt{2}}{3} I_0 \, , 
\nonumber \\
\bra 
\psi_{000} \chi^S _{-3/2} |
\sqrt{2} \sigma_-
\, e^{i \vec q_{eff} \cdot \vec x}\,
| \psi_{000} \chi^\lambda_{-1/2} \ket
&=&
- \sqrt{\frac{2}{3}} I_0 \, , 
\ee
where two independent matrix elements for $\Sigma(3/2^+)$ are shown.  

\subsection{$N(1/2^+) \to$ $p$-wave  baryons}

Let us first consider the transition to $\Lambda(1/2^-)$.  
The rerelvant matrix element is given as 
\be
\bra 
[\psi_{01},  \chi^\rho ]^{1/2}_{-1/2}
|
\sqrt{2} \sigma_-
\, e^{i \vec q_{eff} \cdot \vec x}\,
| \psi_{000} \chi^\rho_{+1/2} 
\ket
= 
\sqrt{\frac{1}{3}} 
\bra 
\chi^\rho _{-1/2} |
\sigma_-
| \chi^\rho_{+1/2} \ket \, 
\bra 
\psi_{010} |\sqrt{2}
\, e^{i \vec q_{eff} \cdot \vec x}\,
| \psi_{000}
\ket \, , 
\ee
where the factor $\sqrt{1/3}$ is the Clebsh-Gordan coefficients in the state 
$[\psi_{01},  \chi^\rho ]^{1/2}_{-1/2}$.  
The radial part is computed as 
\be
\bra 
\psi_{010} | \sqrt{2}
\, e^{i \vec q_{eff} \cdot \vec x}\,
| \psi_{000}\ket
=
\frac{(\alpha^{\prime } \alpha)^{3/2} \alpha^\prime q_{eff}}{A^5}
\, 
e^{-q_{eff}^2/(4A^2)}
\equiv
I_1
\label{I_1}
\ee
and so 
\be
\Lambda(1/2^-): & & 
\bra 
[\psi_{01},  \chi^\rho ]^{1/2}_{-1/2}
|
\sqrt{2} \sigma_-
\, e^{i \vec q_{eff} \cdot \vec x}\,
| \psi_{000} \chi^\rho_{+1/2} \ket
= \sqrt{\frac{1}{3}}\, I_1  \, .
\ee
Other matrix elements can be computed similarly:
\be
\Lambda(3/2^-):& & 
\bra 
[\psi_{01},  \chi^\rho ]^{3/2}_{-1/2}
|
\sqrt{2} \sigma_-
\, e^{i \vec q_{eff} \cdot \vec x}\,
| \psi_{000} \chi^\rho_{+1/2} \ket
=
\sqrt{\frac{2}{3}} I_1 \, , 
\nonumber \\
& & 
\bra 
[\psi_{01},  \chi^\rho ]^{3/2}_{-3/2}
|
\sqrt{2} \sigma_-
\, e^{i \vec q_{eff} \cdot \vec x}\,
| \psi_{000} \chi^\rho_{-1/2} \ket
=
0 \, , 
\nonumber \\
\Sigma(1/2^-):& & 
\bra 
[\psi_{01},  \chi^\lambda ]^{3/2}_{-1/2}
|
\sqrt{2} \sigma_-
\, e^{i \vec q_{eff} \cdot \vec x}\,
| \psi_{000} \chi^\lambda_{+1/2} \ket
=
\frac{1}{3\sqrt{3}} I_1 \, , 
\nonumber \\
\Sigma(3/2^-): & & 
\bra 
[\psi_{01},  \chi^\lambda ]^{3/2}_{-1/2}
|
\sqrt{2} \sigma_-
\, e^{i \vec q_{eff} \cdot \vec x}\,
| \psi_{000} \chi^\lambda_{+1/2} \ket
=
- \frac{1}{3}\sqrt{\frac{2}{3}} I_1 \, , 
\nonumber \\
& & 
\bra 
[\psi_{01},  \chi^\lambda ]^{3/2}_{-3/2}
|
\sqrt{2} \sigma_-
\, e^{i \vec q_{eff} \cdot \vec x}\,
| \psi_{000} \chi^\lambda_{-1/2} \ket
=
0 \, , 
\nonumber \\
\Sigma^\prime(1/2^-): & & 
\bra 
[\psi_{01},  \chi^S ]^{1/2}_{-1/2}
|
\sqrt{2} \sigma_-
\, e^{i \vec q_{eff} \cdot \vec x}\,
| \psi_{000} \chi^\lambda_{+1/2} \ket
=
- \frac{1}{3}\sqrt{\frac{2}{3}} I_1 \, , 
\nonumber \\
\Sigma^\prime(3/2^-): & & 
\bra 
[\psi_{01},  \chi^S ]^{3/2}_{-1/2}
|
\sqrt{2} \sigma_-
\, e^{i \vec q_{eff} \cdot \vec x}\,
| \psi_{000} \chi^\lambda_{+1/2} \ket
=
\frac{1}{3} \sqrt{\frac{2}{15}} I_1 \, , 
\nonumber \\
& & 
\bra 
[\psi_{01},  \chi^S ]^{3/2}_{-3/2}
|
\sqrt{2} \sigma_-
\, e^{i \vec q_{eff} \cdot \vec x}\,
| \psi_{000} \chi^\lambda_{-1/2} \ket
=
\sqrt{\frac{2}{5}} I_1 \, , 
\nonumber \\
\Sigma^\prime(5/2^-): & & 
\bra 
[\psi_{01},  \chi^S ]^{5/2}_{-1/2}
|
\sqrt{2} \sigma_-
\, e^{i \vec q_{eff} \cdot \vec x}\,
| \psi_{000} \chi^\lambda_{+1/2} \ket
=
- \sqrt{\frac{2}{15}} I_1 \, , 
\nonumber \\
& & 
\bra 
[\psi_{01},  \chi^S ]^{5/2}_{-3/2}
|
\sqrt{2} \sigma_-
\, e^{i \vec q_{eff} \cdot \vec x}\,
| \psi_{000} \chi^\lambda_{-1/2} \ket
=
- \sqrt{\frac{4}{15}} I_1 \, .
\ee

\subsection{$N(1/2^+) \to$ $d$-wave baryons}
Computations go in completely similar manner as before, except for the 
radial matrix element
\be
\bra \psi_{020} | \sqrt{2} \, e^{i \vec q_{eff} \cdot \vec x}\,
|\psi_{000} \ket 
= 
\frac{1}{2} \sqrt{\frac{2}{3}} 
\frac{(\alpha \alpha^\prime)^{3/2}}{A^3} 
\left(
\frac{\alpha^\prime q}{A^2}
\right)^2
e^{-q_{eff}^2/(4A^2)}
\equiv 
I_2 \, .
\label{I_2}
\ee
The results are 
\be
\Lambda(3/2^+):
& & 
\bra 
[\psi_{02},  \chi^\rho ]^{3/2}_{-1/2}
|
\sqrt{2} \sigma_-
\, e^{i \vec q_{eff} \cdot \vec x}\,
| \psi_{000} \chi^\rho_{+1/2} \ket
=
- \sqrt{\frac{2}{5}} I_2 \, , 
\nonumber \\
& & 
\bra 
[\psi_{02},  \chi^\rho ]^{3/2}_{-3/2}
|
\sqrt{2} \sigma_-
\, e^{i \vec q_{eff} \cdot \vec x}\,
| \psi_{000} \chi^\rho_{-1/2} \ket
= 0 \, , 
\nonumber \\
\Lambda(5/2^+): 
& & 
\bra 
[\psi_{02},  \chi^\rho ]^{5/2}_{-1/2}
|
\sqrt{2} \sigma_-
\, e^{i \vec q_{eff} \cdot \vec x}\,
| \psi_{000} \chi^\rho_{+1/2} \ket
=
\sqrt{\frac{3}{5}} I_2 \, , 
\nonumber \\
& & 
\bra 
[\psi_{02},  \chi^\rho ]^{5/2}_{-3/2}
|
\sqrt{2} \sigma_-
\, e^{i \vec q_{eff} \cdot \vec x}\,
| \psi_{000} \chi^\rho_{-1/2} \ket
= 0 \, , 
\nonumber \\
\Sigma(3/2^+): 
& & 
\bra 
[\psi_{02},  \chi^\lambda ]^{3/2}_{-1/2}
|
\sqrt{2} \sigma_-
\, e^{i \vec q_{eff} \cdot \vec x}\,
| \psi_{000} \chi^\lambda_{+1/2} \ket
=
\sqrt{\frac{3}{5}} I_2 \, , 
\nonumber \\
& & 
\bra 
[\psi_{02},  \chi^\lambda ]^{3/2}_{-3/2}
|
\sqrt{2} \sigma_-
\, e^{i \vec q_{eff} \cdot \vec x}\,
| \psi_{000} \chi^\lambda_{-1/2} \ket
= 0 \, , 
\nonumber \\
\Sigma(5/2^+): 
& & 
\bra 
[\psi_{02},  \chi^\lambda ]^{5/2}_{-1/2}
|
\sqrt{2} \sigma_-
\, e^{i \vec q_{eff} \cdot \vec x}\,
| \psi_{000} \chi^\lambda_{+1/2} \ket
=
\sqrt{\frac{3}{5}} I_2 \, , 
\nonumber \\
& & 
\bra 
[\psi_{02},  \chi^\lambda ]^{5/2}_{-3/2}
|
\sqrt{2} \sigma_-
\, e^{i \vec q_{eff} \cdot \vec x}\,
| \psi_{000} \chi^\lambda_{-1/2} \ket
= 0 \, , 
\nonumber \\
\Sigma^\prime(1/2^+): 
& & 
\bra 
[\psi_{02},  \chi^S ]^{1/2}_{-1/2}
|
\sqrt{2} \sigma_-
\, e^{i \vec q_{eff} \cdot \vec x}\,
| \psi_{000} \chi^\lambda_{+1/2} \ket
=
\sqrt{\frac{3}{5}} I_2 \, , 
\nonumber \\
\Sigma^\prime(3/2^+): 
& & 
\bra 
[\psi_{02},  \chi^S ]^{3/2}_{-1/2}
|
\sqrt{2} \sigma_-
\, e^{i \vec q_{eff} \cdot \vec x}\,
| \psi_{000} \chi^\lambda_{+1/2} \ket
=
\sqrt{\frac{3}{5}} I_2 \, , 
\nonumber \\
& & 
\bra 
[\psi_{02},  \chi^S ]^{3/2}_{-3/2}
|
\sqrt{2} \sigma_-
\, e^{i \vec q_{eff} \cdot \vec x}\,
| \psi_{000} \chi^\lambda_{-1/2} \ket
= 0 \, , 
\nonumber \\
\Sigma^\prime(5/2^+): 
& & 
\bra 
[\psi_{02},  \chi^S ]^{5/2}_{-1/2}
|
\sqrt{2} \sigma_-
\, e^{i \vec q_{eff} \cdot \vec x}\,
| \psi_{000} \chi^\lambda_{+1/2} \ket
=
\sqrt{\frac{3}{5}} I_2 \, , 
\nonumber \\
& & 
\bra 
[\psi_{02},  \chi^S ]^{5/2}_{-3/2}
|
\sqrt{2} \sigma_-
\, e^{i \vec q_{eff} \cdot \vec x}\,
| \psi_{000} \chi^\lambda_{-1/2} \ket
= 0 \, , 
\nonumber \\
\Sigma^\prime(7/2^+): 
& & 
\bra 
[\psi_{02},  \chi^S ]^{7/2}_{-1/2}
|
\sqrt{2} \sigma_-
\, e^{i \vec q_{eff} \cdot \vec x}\,
| \psi_{000} \chi^\lambda_{+1/2} \ket
=
\sqrt{\frac{3}{5}} I_2 \, , 
\nonumber \\
& & 
\bra 
[\psi_{02},  \chi^S ]^{7/2}_{-3/2}
|
\sqrt{2} \sigma_-
\, e^{i \vec q_{eff} \cdot \vec x}\,
| \psi_{000} \chi^\lambda_{-1/2} \ket
= 0 \, .
\ee

\section{Baryon wave functions}

We summarize the baryon wave functions used in the present calculations~\cite{Hosaka2001}.  
They are constructed by a quark and a diquark, and are expressed as products of 
isospin, spin and orbital wave functions.  
Here we show explicitly spin and orbital parts.  
For orbital wave functions, we employ harmonic oscillator functions as given in appendix C.  

For spin wave functions, using the notation for angular momentum coupling 
$[L_1, L_2]^{L_{\rm tot}}$ 
we employ the three functions
\be
\chi^\rho _m &=& [d^0, \chi]^{1/2}_{m} \, , 
\nonumber \\
\chi^\lambda_m &=& [d^1, \chi]^{1/2}_{m} \, , 
\nonumber \\
\chi^S_m &=& [d^1, \chi]^{3/2}_{m}  \, .
\ee
where $d^S$ denotes the diquark spin function,  and $\chi$ the two component spinor 
for a single quark.  
For the ground baryons we have three states
\be
\Lambda(1/2^+,m) &=&
\psi_{000}(\vec x) \chi^\rho _m \, , 
\nonumber \\
\Sigma(1/2^+,m) &=&
\psi_{000}(\vec x) \chi^\lambda _m \, , 
\nonumber \\
\Sigma(3/2^+,m) &=&
\psi_{000}(\vec x) \chi^S _m \, .
\ee
For the first excited states of negative parity there are seven states 
($\psi_{nlm} \to \psi_{nl} = \psi_{01}$)
\be
\Lambda(1/2^-,m) &=&
[\psi_{01}(\vec x),  \chi^\rho ]^{1/2}_m \, , 
\nonumber \\
\Lambda(3/2^-,m) &=&
[\psi_{01}(\vec x),  \chi^\rho ]^{3/2}_m \, , 
\nonumber \\
\Sigma(1/2^-,m) &=&
[\psi_{01}(\vec x),  \chi^\lambda ]^{1/2}_m \, , 
\nonumber \\
\Sigma(3/2^-,m) &=&
[\psi_{01}(\vec x),  \chi^\lambda ]^{3/2}_m \, , 
\nonumber \\
\Sigma^\prime(1/2^-,m) &=&
[\psi_{01}(\vec x),  \chi^S ]^{1/2}_m \, , 
\nonumber \\
\Sigma^\prime(3/2^-,m) &=&
[\psi_{01}(\vec x),  \chi^S ]^{3/2}_m \, , 
\nonumber \\
\Sigma^\prime(5/2^-,m) &=&
[\psi_{01}(\vec x),  \chi^S ]^{5/2}_m \, .
\ee
Similarly, we obtain the wave functions for the $l=2$ excited baryons.  

Finally, the nucleon wave function is given as 
\be
N = \psi_{000} \frac{1}{\sqrt{2}} 
\left(  \chi^\rho \phi^\rho + \chi^\lambda \phi^\lambda \right) \, , 
\ee
where $\phi^\rho$ and $\phi^\lambda$ are the ispsoin 1/2 wave functions
of the nucleon with three quarks.  

\section{Harmonic oscillator wave functions}

We summarize some of the harmonic oscillator wave functions for low lying states.  
Including the angular and radial parts, they are given as 
\be
\psi_{nlm}(\vec x) = Y_{lm}(\hat x)  R_{nl}(r) \, , 
\ee
where $ R_{nl}(r)$ are 
\be
R_{00}(r) &=& 
\frac{\alpha^{3/2}} {\pi^{1/4}} 2 e^{-(\alpha^{2}/2) r^2} \, , 
\nonumber \\
R_{01}(r) &=& 
\frac{\alpha^{3/2}} {\pi^{1/4}} 
\left(\frac{8}{3}\right)^{1/2} \alpha r e^{-(\alpha^{2}/2) r^2} \, , 
\nonumber \\
R_{10}(r) &=& 
\frac{\alpha^{3/2}} {\pi^{1/4}} 
(2\cdot 3)^{1/2} 
\left(
1 - \frac{2}{3} (\alpha r)^2 
\right)
e^{-(\alpha^{2}/2) r^2} \, , 
\nonumber \\
R_{02}(r) &=& 
\frac{\alpha^{3/2}} {\pi^{1/4}} 
\left(\frac{16}{5 \cdot 3}\right)^{1/2} 
(\alpha r)^2
e^{-(\alpha^{2}/2) r^2} \, .
\ee
The oscillator parameter $\alpha$ is related to the frequency 
$\omega$ by
\be
\alpha = \sqrt{m \omega} = (km)^{1/4} \, , 
\ee
where $k$ is the spring constant. 

\baselineskip 5mm

\end{document}